\documentclass[twocolumn,english,aps,prb,showpacs,amsmath,amssymb,superscriptaddress]{revtex4}
\usepackage[latin9]{inputenc}
\usepackage{color}
\usepackage{amstext}
\usepackage{amssymb}
\usepackage{graphicx}
\usepackage{pdfpages}

\makeatletter
\@ifundefined{textcolor}{}
{%
 \definecolor{BLACK}{gray}{0}
 \definecolor{WHITE}{gray}{1}
 \definecolor{RED}{rgb}{1,0,0}
 \definecolor{GREEN}{rgb}{0,1,0}
 \definecolor{BLUE}{rgb}{0,0,1}
 \definecolor{CYAN}{cmyk}{1,0,0,0}
 \definecolor{MAGENTA}{cmyk}{0,1,0,0}
 \definecolor{YELLOW}{cmyk}{0,0,1,0}
}

\usepackage{color}

\newcommand{\beq}{\begin{eqnarray}}
\newcommand{\eeq}{\end{eqnarray}}

\newcommand{\ys}[1]{\textcolor{black}{#1}}
\newcommand{\ysb}[1]{\textcolor{black}{#1}}

\makeatother

\usepackage{babel}
\begin{document}

\title{Spin anisotropy due to spin-orbit coupling in optimally hole-doped
Ba$_{0.67}$K$_{0.33}$Fe$_{2}$As$_{2}$}

\author{Yu Song}

\email{Yu.Song@rice.edu}

\selectlanguage{english}%

\affiliation{Department of Physics and Astronomy, Rice University, Houston, Texas
77005, USA}

\author{Haoran Man}

\affiliation{Department of Physics and Astronomy, Rice University, Houston, Texas
77005, USA}

\author{Rui Zhang}

\affiliation{Department of Physics and Astronomy, Rice University, Houston, Texas
77005, USA}

\author{Xingye Lu}

\affiliation{Department of Physics and Astronomy, Rice University, Houston, Texas
77005, USA}

\author{Chenglin Zhang}

\affiliation{Department of Physics and Astronomy, Rice University, Houston, Texas
77005, USA}

\author{Meng Wang}

\affiliation{Department of Physics, University of California, Berkeley, California
94720, USA}

\author{Guotai Tan}

\affiliation{Center for Advanced Quantum Studies and Department of Physics, Beijing
Normal University, Beijing 100875, China}

\author{L.-P. Regnault}

\affiliation{Universit\'{e} Grenoble Alpes, 38042 Grenoble, France}

\affiliation{CEA-Grenoble, INAC-MEM-MDN, 38054 Grenoble, France}

\author{Yixi Su}

\affiliation{J{ü}lich Centre for Neutron Science, Forschungszentrum J{ü}lich
GmbH, Outstation at MLZ, D-85747 Garching, Germany}

\author{Jian Kang}

\affiliation{School of Physics and Astronomy, University of Minnesota, Minneapolis,
Minnesota 55455, USA}

\author{Rafael M. Fernandes}

\affiliation{School of Physics and Astronomy, University of Minnesota, Minneapolis,
Minnesota 55455, USA}

\author{Pengcheng Dai}

\email{pdai@rice.edu}

\selectlanguage{english}%

\affiliation{Department of Physics and Astronomy, Rice University, Houston, Texas
77005, USA}

\affiliation{Center for Advanced Quantum Studies and Department of Physics, Beijing
Normal University, Beijing 100875, China}
\begin{abstract}
We use polarized inelastic neutron scattering to study the temperature
and energy dependence of spin space anisotropies in the optimally
hole-doped iron pnictide Ba$_{0.67}$K$_{0.33}$Fe$_{2}$As$_{2}$
($T_{{\rm c}}=38$ K). In the superconducting state, while the high-energy
part of the magnetic spectrum is nearly isotropic, the low-energy
part displays a pronouced anisotropy, manifested by a $c$-axis polarized
resonance. We also observe that the spin anisotropy in superconducting
Ba$_{0.67}$K$_{0.33}$Fe$_{2}$As$_{2}$ extends to higher energies
compared to electron-doped BaFe$_{2-x}TM_{x}$As$_{2}$ ($TM=$Co,
Ni) and isovalent-doped BaFe$_{2}$As$_{1.4}$P$_{0.6}$, suggesting
a connection between $T_{\rm c}$ and the energy scale of the spin anisotropy.
In the normal state, the low-energy spin anisotropy for optimally
hole- and electron-doped iron pnictides onset at temperatures similar
to the temperatures at which the elastoresistance deviate from Curie-Weiss
behavior, pointing to a possible connection between the two phenomena.
Our results highlight the relevance of the spin-orbit coupling to
the superconductivity of the iron pnictides. 
\end{abstract}


\maketitle

\section{Introduction}
The parent compounds of iron pnictide superconductors, such as LaFeAsO
and BaFe$_{2}$As$_{2}$, form stripe antiferromagnetic (AF) order
at $T_{{\rm N}}$ below a tetragonal-to-orthorhombic structural transition
temperature $T_{{\rm S}}$ {[}inset in Fig. 1(b){]} \cite{kamihara,cruz,johnston,dai}.
Superconductivity can be induced by partially replacing Ba by K in
BaFe$_{2}$As$_{2}$ to form hole-doped Ba$_{1-x}$K$_{x}$Fe$_{2}$As$_{2}$
\cite{Rotter,Avci,Wasser,Allred,Allred2} or by partially replacing
Fe by $TM$ ($TM=$Co, Ni) to form electron-doped BaFe$_{2-x}TM_{x}$As$_{2}$
\cite{ASefat,LJLi09,DKPratt2011,XYLuPRL}. Importantly, the resulting
phase diagrams exhibit significant asymmetry between electron- and
hole-doping {[}Figs. 1(a) and 1(b){]} \cite{johnston,dai}. For instance,
while near optimal doping the stripe AF order becomes incommensurate
for electron-doped BaFe$_{2-x}TM_{x}$As$_{2}$ \cite{DKPratt2011,XYLuPRL}{[}see arrow
in Fig. 1(b){]}, a double-\textbf{Q} tetragonal magnetic structure
with ordered moments along the $c$-axis is observed in hole-doped
Ba$_{1-x}$K$_{x}$Fe$_{2}$As$_{2}$ {[}see region of the phase diagram near the arrow in Fig. 1(a){]} \cite{Avci,Wasser,Allred,Allred2}.

Nevertheless, upon entering the superconducting state, a magnetic
resonance mode appears in the magnetic spectrum in both cases at the AF wave vector (${\bf Q}_{{\rm AF}}$) \cite{Christianson2008,CZhang2011,SChi2009,MDLumsden2009}.
Furthermore, by measuring the splitting of the electronic states at
high-symmetry points in reciprocal space \cite{Fernandes14}, angle-resolved
photoemission spectroscopy (ARPES) measurements find that spin-orbit
coupling (SOC) is present in both electron- and hole-doped iron pnictides
with a similar energy scale $\sim10$ meV \cite{Borisenko}. Also
common to both optimally electron-doped BaFe$_{2-x}TM_{x}$As$_{2}$
\cite{JHChu2010,XYLu2016} and hole-doped Ba$_{1-x}$K$_{x}$Fe$_{2}$As$_{2}$
\cite{JJYing,Blomberg} is the presence of electronic nematic fluctuations,
as revealed by the elastoresistance -- i.e. the rate of change of
the resistivity anisotropy with respect to applied in-plane uniaxial
strain {[}Fig. 1(c){]} \cite{Fernandes2011}. The elastoresistance
diverges with a Curie-Weiss form for both classes of materials as
well as for isovalent-doped BaFe$_{2}$As$_{1.4}$P$_{0.6}$ \cite{Kuo2016}.
Deviation from the Curie-Weiss behavior is seen in both optimally
electron- and hole-doped BaFe$_{2}$As$_{2}$ at low temperatures,
while no deviation is seen in BaFe$_{2}$As$_{1.4}$P$_{0.6}$ down
to $T_{{\rm c}}$ {[}Fig. 1(d){]} \cite{Kuo2016}.

In addition to its impact on the electronic spectrum \cite{Borisenko,PJohnson},
SOC also converts crystalline anisotropies into anisotropies in spin
space, as seen from nuclear magnetic resonance studies \cite{ZLi}. The spin anisotropy resulting from SOC plays an essential role
for the double-${\bf Q}$ magnetic phase \cite{Avci,Wasser,Allred,Allred2},
in which the ordered moments align along the $c$-axis \cite{MChristensen}.
If SOC was absent, the spin excitations in the paramagnetic tetragonal
state of the iron pnictides would be isotropic in spin space {[}Fig.
1(e){]}. However, due to the presence of a sizable SOC, an anisotropy
is developed in the spin excitations,
which can be quantitatively determined by neutron polarization analysis
\cite{Moon}. In the antiferromagnetically ordered phases of the parent
compounds BaFe$_{2}$As$_{2}$ and NaFeAs \cite{qhunag,SLLi}, where
the ordered moments point parallel to the orthorhombic $a$-axis {[}inset
in Fig. 1(b){]}, spin waves exhibit significant anisotropy, with $c$-axis
polarized spin waves occurring at lower energy compared to $b$-axis
polarized spin waves \cite{NQureshi2012,CWangPRX,YSong2013}. 
To elucidate the relevance of SOC to superconductivity, it is instructive
to compare the behavior of the spin anisotropy in hole-doped and electron-doped
BaFe$_{2}$As$_{2}$, since the maximum values of $T_{\rm c}$ are quite
different in these two cases -- $T_{\rm c}\approx38$ K for optimally
hole-doped Ba$_{0.67}$K$_{0.33}$Fe$_{2}$As$_{2}$ and $T_{\rm c}\approx25$
K for optimally electron-doped BaFe$_{1.86}$Co$_{0.14}$As$_{2}$.
Previous analysis of the electron-doped case revealed that the spin
anisotropy persists in the paramagnetic tetragonal phase for doping
levels up to or slightly beyond optimal doping \cite{Lipscombe,HQLuo2013,PSteffens,CZhang2013,NQureshi2014,CZhang2014},
but vanishes in the \ys{well-}overdoped regime \cite{MSLiu2012,CZhang2014}. 

In this paper, we present polarized neutron scattering studies of
spin excitations in optimally hole-doped Ba$_{0.67}$K$_{0.33}$Fe$_{2}$As$_{2}$ \cite{CZhang2013,CZhang2011}. 
In the normal
state, \ys{we find that} the spin anisotropy of Ba$_{0.67}$K$_{0.33}$Fe$_{2}$As$_{2}$
persists to $\sim100$ K for $E=3$ meV, similarly to the case of
near-optimally electron-doped BaFe$_{2-x}TM_{x}$As$_{2}$, where
spin anisotropy at ${\bf Q}_{{\rm AF}}=(1,0,1)$ was found below $E\approx7$
meV and up to $\sim70$ K \cite{HQLuo2013,PSteffens}. \ys{We associate the onset of normal state spin anisotropy with the nematic susceptibility deviating from Curie-Weiss behavior measured via elastoresistance {[}see vertical arrows in Fig. 1(d){]} \cite{Kuo2016}, indicating an important role of spin excitations in transport properties of iron pnictides.}

Upon entering the superconducting state, \ys{we find that} while
at high energies \ys{($E\geq 14$ meV)} the spectrum is nearly isotropic \ys{as found in previous work \cite{CZhang2013}}, at low energies
the resonance mode is strongly anisotropic, being dominated by a $c$-axis
polarized component. We attribute this behavior to the fact that the
superconducting state is close to the double-${\bf Q}$ magnetic phase,
in which the magnetic moments point out-of-plane \cite{Wasser,Allred}.
Indeed, by adding a spin-anisotropic term that favors $c$-axis spin
orientation in a simple two-band theoretical model, we find that the
resonance mode in the $c$-axis polarized channel has in general a
lower energy than in other channels, and that this energy difference
increases as the magnetically ordered state is approached. Our analysis
also reveals an interesting correlation between the energy scale of
the spin anisotropy in the superconducting state and
$T_{\rm c}$ \cite{Lipscombe,HQLuo2013,PSteffens,CZhang2013,NQureshi2014,CZhang2014,DHu},
suggesting that SOC is an integral part of the superconductivity of
iron pnictides.

\begin{figure}
\includegraphics[scale=0.5]{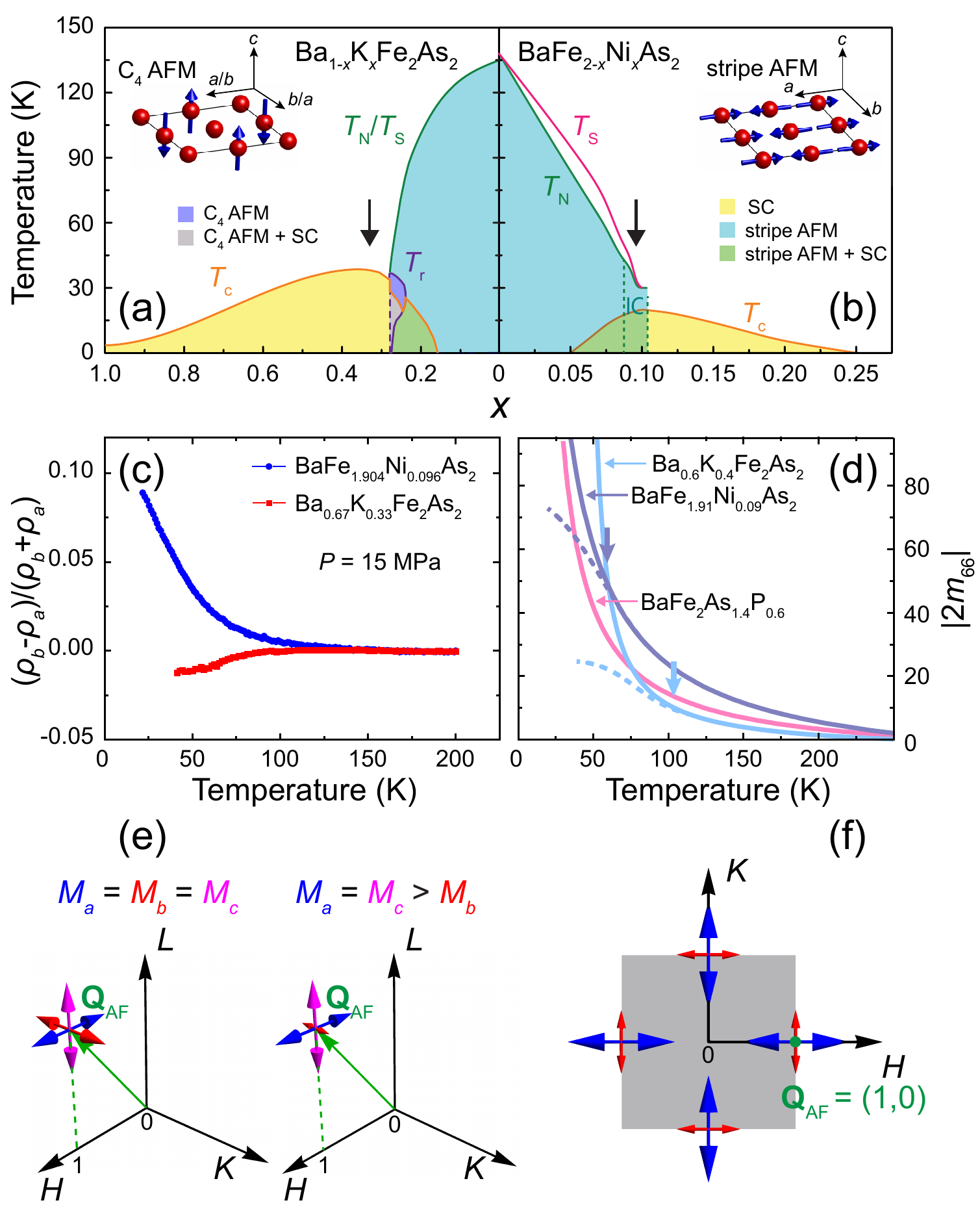} \protect\caption{ (Color online) The electronic phase diagrams of (a) hole- and (b)
electron-doped BaFe$_{2}$As$_{2}$. While parent compounds of iron
pnictides have stripe AF order {[}inset in (b){]} \cite{cruz}, the
tetragonal double-\textbf{Q} $C_{4}$ AF order is found in hole-doped
BaFe$_{2}$As$_{2}$ near optimal superconductivity {[}inset in (a){]}
\cite{Avci,Wasser,Allred,Allred2}. $T_{{\rm S}}$, $T_{{\rm N}}$,
$T_{{\rm c}}$ and $T_{{\rm r}}$ mark the tetragonal-to-orthorhombic
structural transition, the paramagnetic-to-AF transition, the superconducting
transition and the transition into the $C_{4}$ magnetic phase. The
phase diagrams in (a) and (b) are adapted from Refs. \cite{Bohmer}
and \cite{dai}. (c) Resistivity anisotropy $(\rho_{a}-\rho_{b})/(\rho_{a}+\rho_{b})$
of BaFe$_{1.904}$Ni$_{0.096}$As$_{2}$ and Ba$_{0.67}$K$_{0.33}$Fe$_{2}$As$_{2}$
under uniaxial pressure of $P=15$ MPa \ys{measured using a 
mechanical clamp that can vary applied pressure \textit{in-situ}} \cite{supplementary}. 
(d) Elastoresistance $|2m_{66}|$
for optimally-doped BaFe$_{1.91}$Ni$_{0.09}$As$_{2}$, Ba$_{0.6}$K$_{0.4}$Fe$_{2}$As$_{2}$,
and BaFe$_{2}$As$_{1.4}$P$_{0.6}$, adapted from Ref. \cite{Kuo2016}.
The solid lines are Curie-Weiss fits to the data and dashed lines
represent deviations from the Curie-Weiss form. Vertical arrows mark
the temperature at which such deviations begin. (e) Schematic of isotropic spin excitations (left) and anisotropic
spin excitations (right), with the sizes of arrows centered at ${\bf Q}_{{\rm AF}}$
representing the intensities of spin excitations polarized along different
directions. (f) \ys{In-plane} spin anisotropy discussed in this work \ys{(represented by red and blue arrows of different sizes)} preserves four-fold
rotational symmetry of the tetragonal unit cell \ysb{because ${\bf Q}_{{\rm AF}}$ is at an edge of the Brillouin zone of the unfolded tetragonal (i.e. 1-Fe) unit cell \cite{JTPark2010}, depicted by the shaded gray area}.}
\end{figure}

\section{Experimental results}

Polarized inelastic neutron scattering measurements were carried out
using the IN22 triple-axis spectrometer at Institut Laue-Langevin,
Grenoble, France. We studied Ba$_{0.67}$K$_{0.33}$Fe$_{2}$As$_{2}$
single crystals ($a=b\approx5.56$ {\AA }, $c=13.29$ {\AA })
co-aligned in the $[H,0,L]$ scattering plane
used in previous works \cite{CZhang2013,MWang2013}. We use the orthorhombic notation suitable for AF ordered iron pnictides
even though Ba$_{0.67}$K$_{0.33}$Fe$_{2}$As$_{2}$
has a tetragonal structure and is paramagnetic at all temperatures
\cite{CZhang2011,CZhang2013}. Thus, the momentum transfer is ${\bf Q}=H{\bf a^{*}}+K{\bf b^{*}}+L{\bf c^{*}}$,
with ${\bf a^{*}}=\frac{2\pi}{a}{\bf \hat{a}}$, ${\bf b^{*}}=\frac{2\pi}{b}{\bf \hat{b}}$
and ${\bf c^{*}}=\frac{2\pi}{c}{\bf \hat{c}}$, where $H$, $K$ and
$L$ are Miller indices. In this notation, magnetic order in BaFe$_{2}$As$_{2}$
occurs at ${\bf Q}_{{\rm AF}}=(1,0,L)$ with $L=1,3,5\dots$ {[}Fig.
1(e){]}. Three neutron spin-flip (SF) cross sections $\sigma_{x}^{{\rm SF}}$,
$\sigma_{y}^{{\rm SF}}$ and $\sigma_{z}^{{\rm SF}}$ were measured,
with the usual convention $x\parallel{\bf Q}$, $y\perp{\bf Q}$ and
in the scattering plane, and $z$ perpendicular to the scattering
plane. Magnetic neutron scattering directly measures the magnetic
scattering function $S^{\alpha\beta}({\bf Q},E)$, which is proportional
to the imaginary part of the dynamic susceptibility through the Bose
factor, $S^{\alpha\beta}({\bf Q},E)\propto[1-\text{exp}(-\frac{E}{k_{{\rm B}}T})]^{-1}\text{Im}\chi^{\alpha\beta}({\bf Q},E)$
\cite{Furrer}. Following earlier works \cite{YSong2013,HQLuo2013,CZhang2013,CZhang2014,MSLiu2012},
we denote the diagonal components of the magnetic scattering function
$S^{\alpha\alpha}$ as $M_{\alpha}$.  
$M_y$ and $M_z$ can be obtained from measured SF cross sections through $\sigma_{x}^{{\rm SF}}-\sigma_{y}^{{\rm SF}}\propto M_{y}$ and $\sigma_{x}^{{\rm SF}}-\sigma_{z}^{{\rm SF}}\propto M_{z}$.
$M_{y}$ and $M_{z}$ are related to $M_{a}=M_{100}$, 
$M_{b}=M_{010}$ and $M_{c}=M_{001}$ through $M_{y}=\sin^{2}\theta M_{a}+\cos^{2}\theta M_{c}$
and $M_{z}=M_{b}$ \cite{YSong2013,HQLuo2013,CZhang2013,CZhang2014,MSLiu2012},
with $\theta$ being the angle between ${\bf Q}$ and ${\bf a^{*}}$/${\bf a}$.
Anisotropy between $M_{a}$ and $M_{b}$ at ${\bf Q}_{{\rm AF}}$
is allowed in the paramagnetic tetragonal state of iron pnictides and does not
break four-fold rotation symmetry of the lattice \ysb{because ${\bf Q}_{{\rm AF}}$ is at an edge of the Brillouin zone of the unfolded tetragonal (1-Fe) unit cell \cite{JTPark2010}}, as depicted in Fig.
1(f). \ysb{Another manifestation of the lack of four-fold rotational symmetry for magnetic excitations at ${\bf Q}_{\rm AF}$ is the anisotropic in-plane correlation lengths seen in the paramagnetic tetragonal states of BaFe$_2$As$_2$ \cite{LWHarriger2011} and CaFe$_2$As$_2$ \cite{SODiallo2010}.} By obtaining $M_{y}$ and $M_{z}$ at two
equivalent wave vectors with different $\theta$, it is then possible
to obtain $M_{a}$, $M_{b}$, and $M_{c}$ \cite{YSong2013,CZhang2014,HQLuo2013,CWangPRX}.

\begin{figure}
\includegraphics[scale=0.55]{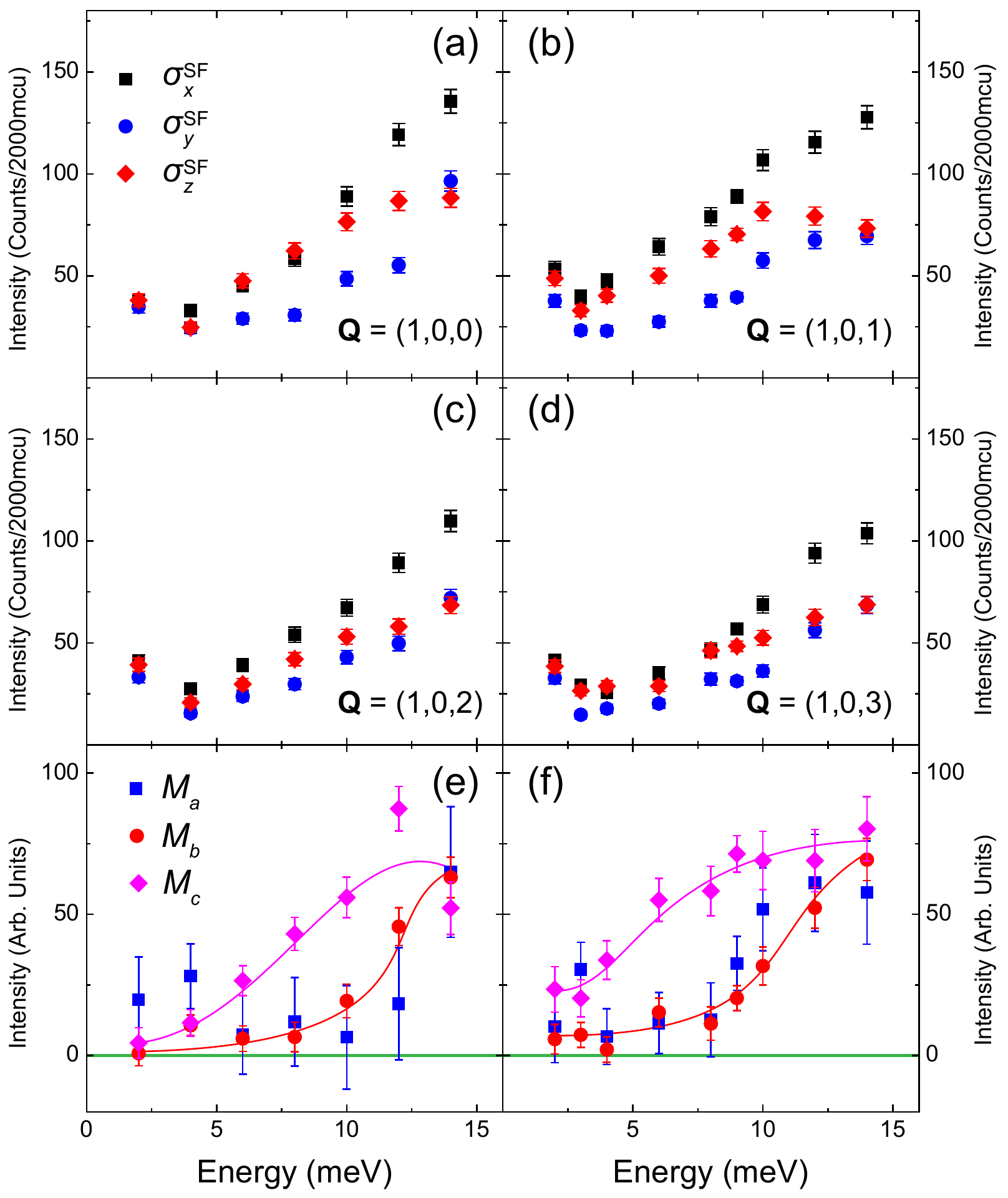} \protect\caption{(Color online) Constant-${\bf Q}$ scans of $\sigma_{x}^{{\rm SF}}$,
$\sigma_{y}^{{\rm SF}}$ and $\sigma_{z}^{{\rm SF}}$ at ${\bf Q}=(1,0,L)$\textcolor{red}{{}
}for (a) $L=0$, (b) $L=1$, (c) $L=2$ and (d) $L=3$ measured at
2 K. Using the measured cross sections in (a) - (d), $M_{a}$, $M_{b}$
and $M_{c}$ are obtained for (e) even and (f) odd $L$. The solid
lines are guides to the eye. }
\end{figure}

Figure 2 summarizes constant-${\bf Q}$ scans at 2 K and ${\bf Q}=(1,0,L)$
with $L$ = 0, 1, 2 and 3. 
From Fig. 2(a)-(d), it is clear that $\sigma_{x}^{{\rm SF}}>\sigma_{z}^{{\rm SF}}\geq\sigma_{y}^{{\rm SF}}$
below $E\approx14$ meV, meaning that spin anisotropy exists below
this energy \ys{while excitations above this energy are isotropic as shown in previous work} \cite{CZhang2013}. Although magnetic order is fully suppressed
in Ba$_{0.67}$K$_{0.33}$Fe$_{2}$As$_{2}$, the spin gap $E_{g}$
in the superconducting state displays strong $L$ dependence, with
$E_{g}\approx0.75$ meV for odd $L$ and $E_{g}\approx5$ meV for
even $L$ \cite{CZhang2011,CZhang2013}. From Figs. 2 (b) and 2(d),
we observe that the small gap for odd $L$ is due to $M_{y}$, with
$\sigma_{x}^{{\rm SF}}\approx\sigma_{z}^{{\rm SF}}>\sigma_{y}^{{\rm SF}}$
for $E\lesssim5$ meV. Magnetic excitations are gapped in the same
energy range for even $L$ as can be seen in Figs. 2(a) and 2(c),
with $\sigma_{x}^{{\rm SF}}\approx\sigma_{y}^{{\rm SF}}\approx\sigma_{z}^{{\rm SF}}$.
$M_{a}$, $M_{b}$ and $M_{c}$ for even and odd $L$ are shown in
Figs. 2 (e) and (f), respectively. While $M_{b}$ is weakly $L$-dependent,
$M_{c}$ clearly displays different behaviors for even and odd $L$.
Because in the energy range $5\lesssim E\lesssim10$ meV $M_{c}$
dominates and is dispersive along $L$, we uniquely identify it with
the the anisotropic resonance that disperses along $L$ which was
previously observed \ys{in the same sample} \cite{CZhang2013}.

\begin{figure}
\includegraphics[scale=0.5]{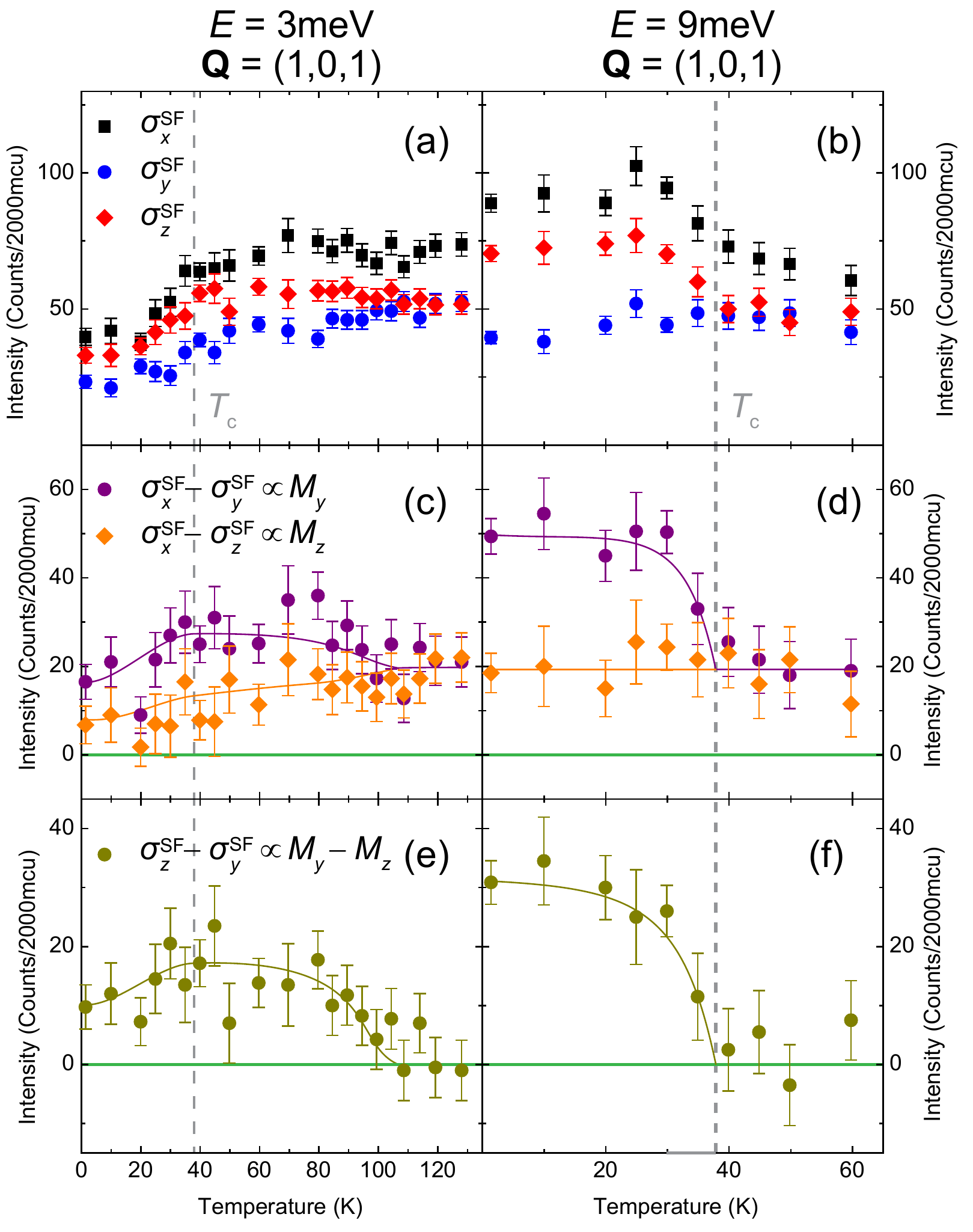} \protect\caption{ (Color online) Temperature scans of $\sigma_{x}^{{\rm SF}}$, $\sigma_{y}^{{\rm SF}}$
and $\sigma_{z}^{{\rm SF}}$ at ${\bf Q}=(1,0,1)$ with (a) $E=3$
meV and (b) $E=9$ meV. The differences $\sigma_{x}^{{\rm SF}}-\sigma_{y}^{{\rm SF}}$
and $\sigma_{x}^{{\rm SF}}-\sigma_{z}^{{\rm SF}}$ which are respectively
proportional to $M_{y}$ and $M_{z}$ are shown for (c) $E=3$ meV
and (d) $E=9$ meV. \ys{The differences $\sigma_{z}^{{\rm SF}}-\sigma_{y}^{{\rm SF}}$ which is
proportional to $M_{y}-M_{z}$ are shown for (e) $E=3$ meV
and (f) $E=9$ meV.} The solid lines are guides to the eye. The dashed
vertical lines mark $T_{{\rm c}}$. }
\end{figure}

To gain further insight into the spin anisotropy of Ba$_{0.67}$K$_{0.33}$Fe$_{2}$As$_{2}$,
we carried out temperature scans at ${\bf Q}_{{\rm AF}}=(1,0,1)$
for $E=3$ meV and $E=9$ meV, as shown in Figs. 3(a) and 3(b). At
$E=3$ meV, the spin anisotropy with $\sigma_{z}^{{\rm SF}}>\sigma_{y}^{{\rm SF}}$
persists up to $\sim100$ K . Although below $T_{{\rm c}}$ the magnetic
signal is suppressed in all three SF cross sections, the normal-state
anisotropy persists {[}Fig. 3(a){]}. At $E=9$ meV, the spin anisotropy
disappears above $T_{{\rm c}}$, suggesting that the main contribution
to the spin anisotropy in the superconducting state arises from the
anisotropic resonance mode \cite{CZhang2013}. In Figs. 3(c) and 3(d),
$\sigma_{x}^{{\rm SF}}-\sigma_{y}^{{\rm SF}}\propto M_{y}$ and $\sigma_{x}^{{\rm SF}}-\sigma_{z}^{{\rm SF}}\propto M_{z}$
are shown. At $E=3$ meV, $M_{y}>M_{z}$ for $T\lesssim100$ K and
both are suppressed below $T_{{\rm c}}$. At $E=9$ meV, while a clear
resonance mode with an order-parameter-like temperature dependence
is seen in $M_{y}$, $M_{z}$ remains constant across $T_{{\rm c}}$.
\ys{The temperature onset of spin anisotropy is more clearly seen in Fig. 3(e) and (f), which plots 
$\sigma_{z}^{{\rm SF}}-\sigma_{y}^{{\rm SF}}\propto M_{y}-M_{z}$ for $E = 3$ meV and $E = 9$ meV, respectively.} 

\begin{figure}
\includegraphics[scale=0.55]{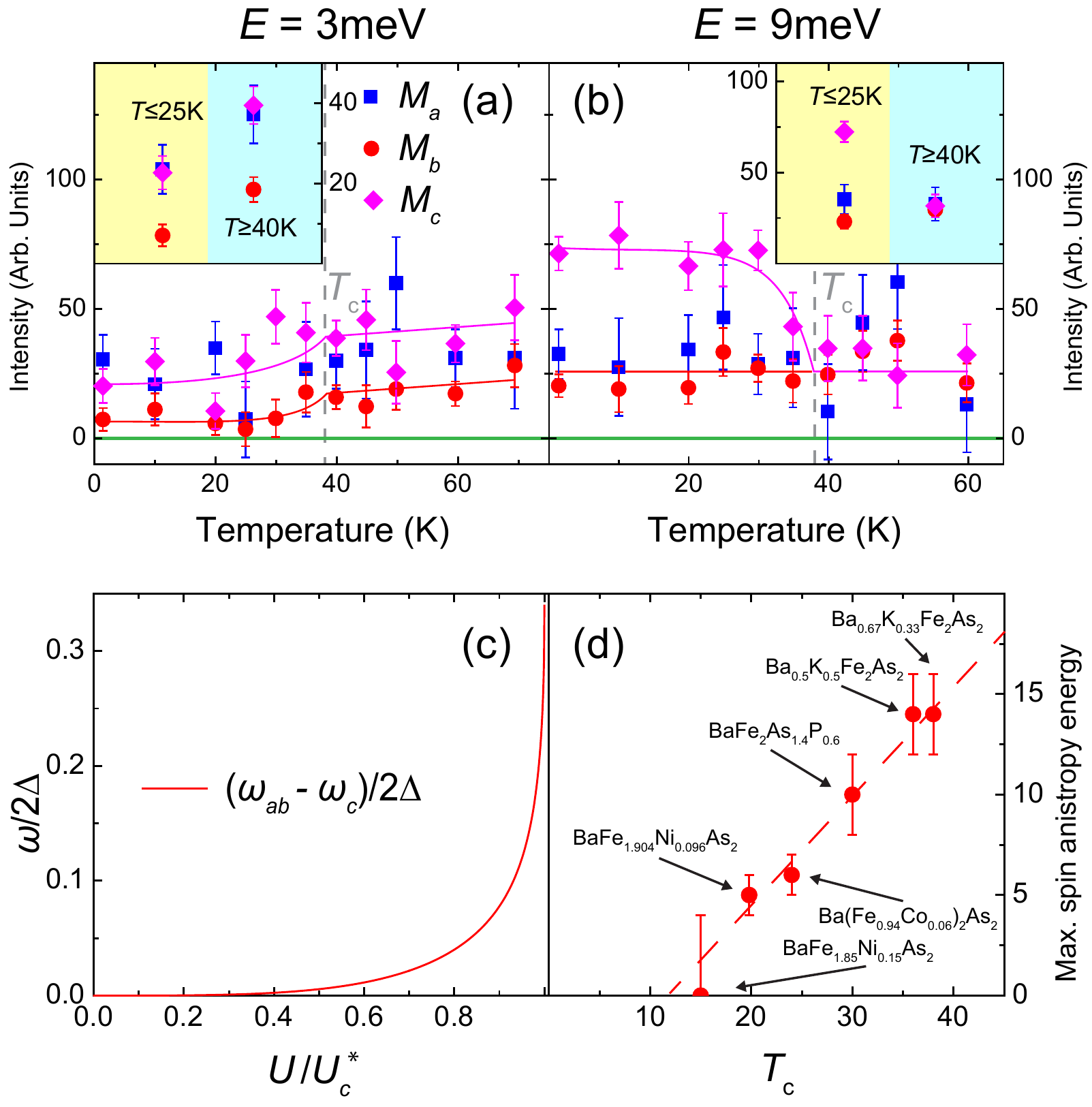} \protect\caption{(Color online) Temperature dependence of $M_{a}$, $M_{b}$ and $M_{c}$
for (a) $E=3$ meV and (b) $E=9$ meV. The solid lines are guides
to the eye and dashed vertical lines mark $T_{{\rm c}}$. The insets
in (a) and (b) show binned $M_{a}$, $M_{b}$ and $M_{c}$ from (a)
and (b) for $T\leq25$ K and $T\geq38$ K. The insets share the same
$y$-axis label as (a) and (b).
(c) Difference between the energies of the $ab$-polarized and the
$c$-polarized resonance modes obtained in our theoretical model.
$U$ is the electronic interaction that triggers long-range magnetic
order with the moments pointing along the $c$-axis when $U=U_{c}^{*}$
\cite{supplementary}. (d) Maximum energy at which
spin anisotropy is observed in the superconducting state of several
doped BaFe$_{2}$As$_{2}$ compounds. Results are obtained from Refs.
\cite{PSteffens,DHu,CZhang2013,NQureshi2014,HQLuo2013,MSLiu2012}.
For overdoped BaFe$_{1.85}$Ni$_{0.15}$As$_{2}$ the excitations
are isotropic but low energy excitations are gapped below $E\approx4$
meV \cite{MSLiu2012}, therefore for this compound we assign the maximum
spin anisotropy energy to be 0, but with an uncertainty of 4 meV. }
\end{figure}

To obtain the temperature dependence of $M_{a}$, $M_{b}$ and $M_{c}$,
we measured $\sigma_{x}^{{\rm SF}}$, $\sigma_{y}^{{\rm SF}}$ and
$\sigma_{z}^{{\rm SF}}$ at ${\bf Q}_{{\rm AF}}=(1,0,3)$ for $E=3$
meV and $E=9$ meV \cite{supplementary}. Combining the temperature
dependence for $L=1$ and $L=3$, $M_{a}$, $M_{b}$ and $M_{c}$
are obtained for odd $L$ as shown in Fig. 4(a) and 4(b). At $E=3$
meV, $M_{a}\approx M_{c}>M_{b}$ within the probed temperature range,
and all three channels decrease in intensity below $T_{{\rm c}}$.
At $E=9$ meV, $M_{a}$ and $M_{b}$ display a weak temperature dependence
while $M_{c}$ is sharply affected by $T_{{\rm c}}$. To corroborate
our conclusion, we binned data points in Fig. 4(a) and 4(b) that are
well below $T_{{\rm c}}$ ($T\leq25$ K) and above $T_{{\rm c}}$
($T\geq40$ K), as shown in the insets of Figs. 4(a) and 4(b). While
magnetic excitations at $E=3$ meV are suppressed upon entering the
superconducting state, the polarization of these magnetic excitations
seems to remain the same, persisting up to $T\approx100$ K. On the
other hand, at $E=9$ meV, magnetic excitations are nearly isotropic
above $T_{{\rm c}}$, while $M_{c}>M_{a}\approx M_{b}$ well below
$T_{{\rm c}}$. Therefore, the
$c$-axis polarized anisotropic resonance is directly coupled to superconductivity
with an order-parameter-like temperature dependence.

\section{Discussion and Conclusion}

To understand the origin of this $c$-axis polarized spin resonance,
we consider a simple two band model \cite{supplementary} in which
the resonance mode arises due to the sign change of the gap function
between a hole pocket and an electron pocket displaced from each other
by the AF ordering vector ${\bf Q}_{{\rm AF}}$ \cite{Korshunov08,Maiti11,Korshunov13}. 
Without SOC, the energy
of the resonance mode is the same for all polarizations, being close
to $2\Delta$ far from the putative magnetic quantum phase transition
inside the superconducting dome [$U\ll U_{c}^{*}$ in Fig. 4(c)],
but vanishing as the transition is approached [$U\rightarrow U_{c}^{*}$
in Fig. 4(c)]. SOC, however, promotes a spin anisotropy term that
makes the magnetic moments point along the $c$-axis for hole-doped
compounds \cite{MChristensen}. As a result, the energy of the resonance
mode polarized along the $c$-axis is suppressed much faster as the
magnetic transition is approached, yielding $\omega_{c}<\omega_{ab}$
{[}Fig. 4(c){]}. This behavior is in qualitative agreement with our
experimental results, with the resonance seen in $M_{c}$ indeed at
lower energies. It should also be noted that our model does not capture the broadening
of the resonance, which is rather pronounced in the experimental data.
Our simple model has two additional consequences: first, as the system
is overdoped and moves farther from the magnetically ordered state,
the resonance mode should become more isotropic. 
\ys{While spin anisotropy persists in slightly overdoped Ba$_{0.5}$K$_{0.5}$Fe$_2$As$_2$ ($T_{\rm c} = 36$ K) \cite{NQureshi2014}, how it evolves in K-well-overdoped} samples remains to be seen. 
Furthermore, because in
electron-doped compounds the moments point along the $a$ direction,
the resonance is expected to be polarized along the $a$-axis. Although
this is the case in electron-doped NaFe$_{0.985}$Co$_{0.015}$As
\cite{CZhang2014}, the sample studied had long-range AF order. For
electron-doped Ba(Fe$_{0.94}$Co$_{0.06}$)$_{2}$As$_{2}$ \cite{PSteffens},
the anisotropic resonance was argued to be also polarized along $c$-axis,
based on the assumption $M_{a}$=$M_{b}$ and the observation $M_{y}>M_{z}=0$
for the anisotropic resonance. As we have shown here and in previous
work \cite{HQLuo2013}, even in the tetragonal state $M_{a}$ and
$M_{b}$ are not necessarily the same and it is unclear whether there
is also significant resonance spectral weight polarized along the
$a$-axis in previous work \cite{PSteffens}. 
\ys{Spin anisotropy of spin excitations has also been detected in the superconducting states of LiFeAs \cite{NQureshi2014b} and FeSe$_{0.5}$Te$_{0.5}$ \cite{PBabkevich2011}, consistent with significant spin-orbit coupling detected by ARPES \cite{Borisenko,PJohnson} in these systems.}  

The normal state spin anisotropy at low energies persists to a temperature
significantly higher than $T_{\rm c}$ {[}$\sim70$ K in BaFe$_{1.094}$Ni$_{0.096}$As$_{2}$
\cite{HQLuo2013} and $\sim100$ K in Ba$_{0.67}$K$_{0.33}$Fe$_{2}$As$_{2}$,
Fig. 3(e){]} for both electron- and hole-doped BaFe$_{2}$As$_{2}$
near optimal doping. The temperature at which spin anisotropy onsets
is similar to the temperature at which the nematic susceptibility
deviates from Curie-Weiss behavior \cite{Kuo2016}, suggesting a common
origin for both phenomena {[}Fig. 1(d){]}. For optimally-doped BaFe$_{2}$As$_{1.4}$P$_{0.6}$,
whose nematic susceptibility shows no deviation from the Curie-Weiss
form {[}Fig. 1(d){]} \cite{Kuo2016}, no spin anisotropy is observed
right above $T_{{\rm c}}$ \cite{DHu}. While disorder is likely to
play an important role in explaining this deviation from Curie-Weiss
behavior in the elastoresistance \cite{Kuo2016}, our results suggest
that the spin anisotropy may also be important. Indeed, previous INS
experiments revealed the intimate relationship between nematicity
and magnetic fluctuations \cite{Lu14,Zhang15,WLZhang2016}. Theoretically, the
nematic susceptibility increases with increasing magnetic fluctuations
in all polarization channels \cite{Fernandes2011}. However, once
a spin anisotropy sets in, fluctuations related to the spin components
perpendicular to the easy axis increase more slowly with decreasing
temperature. As a result, the nematic susceptibility should also increase
more slowly, which may contribute to the deviation from Curie-Weiss
behavior observed experimentally.

Finally, the maximum energies at which spin anisotropy is observed
in the superconducting states of several doped BaFe$_{2}$As$_{2}$
compounds are plotted as function of $T_{{\rm c}}$ in Fig. 4(d).
Note that the spin anisotropy of the resonance in the superconducting
state is also present in BaFe$_{2}$As$_{1.4}$P$_{0.6}$, despite
the absence of spin anisotropy in the normal state \cite{DHu}. We
note a clear positive correlation between the energy scale of the
spin anisotropy and $T_{{\rm c}}$, suggesting SOC to be an important
ingredient for understanding superconductivity in iron pnictides.

\section{Acknowledgments}
We thank Ilya Eremin, Qimiao Si, and Jiangping Hu for useful discussions.
The neutron work at Rice is supported by the U.S. NSF-DMR-1362219
and DMR-1436006 (P.D.). This work is also supported by the Robert
A. Welch Foundation Grant Nos. C-1839 (P.D.). Work performed by R.M.F.
and J.K. is supported by the U.S. Department of Energy, Office of
Science, Basic Energy Sciences, under Award number DE-SC0012336.



\section*{Supplementary material}

# Supplementary Information:

### Determination of $M_a$, $M_b$ and $M_c$

Using the method described in the supplementary of Ref. [1] and data for $\mathbf{Q} = (1,0,L)$ with $L = 1$ and $3$ ($L = 0$ and $2$), we extracted $M_a$, $M_b$ and $M_c$ for the magnetic zone center (zone boundary along $L$) with odd (even) $L$. The data with $L = 3$ used in combination with the $L = 1$ data [Fig. 3(a)-(b)] to obtain $M_a$, $M_b$ and $M_c$ in Fig. 4(a)-(b) are shown here in Supplementary Fig. 1. In the analysis we ignored the differences of sample illumination volume and convolution with instrumental resolution between $L = 1$ and $3$, and set the scale factor that accounts for these differences to be $r = 1$ [1]. In previous works [1–4] $r$ is found to be close to unity in all cases and our results are qualitatively robust when $r$ deviates slightly from 1.

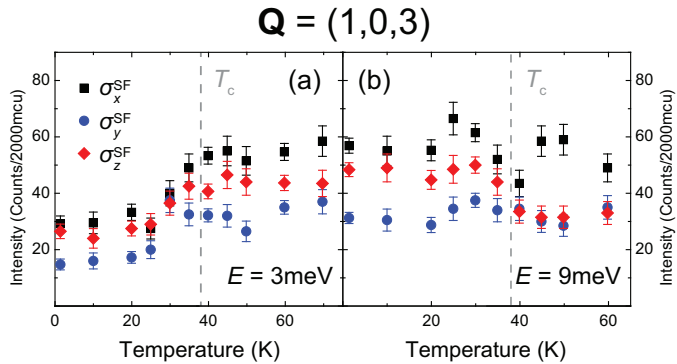

Supplementary Figure 1: Temperature scans of $\sigma_x^{\mathrm{SF}}$, $\sigma_y^{\mathrm{SF}}$ and $\sigma_z^{\mathrm{SF}}$ at $\mathbf{Q} = (1,0,3)$ with (a) $E = 3$ meV and (b) $E = 9$ meV.

### Measurement of resistivity anisotropy under uniaxial pressure

Previous measurements of resistivity anisotropy in $\mathrm{Ba}_{1-x}\mathrm{K}_x\mathrm{Fe}_2\mathrm{As}_2$ demonstrated that compared to their electron-doped counterparts, the hole-doped compounds have much smaller and reversed resistivity anisotropy [5, 6]. Near optimal doping resistivity anisotropy was found to disappear in $\mathrm{Ba}_{0.66}\mathrm{K}_{0.34}\mathrm{Fe}_2\mathrm{As}_2$ [6], however recent measurements revealed significant elastoresistance in $\mathrm{Ba}_{0.6}\mathrm{K}_{0.4}\mathrm{Fe}_2\mathrm{As}_2$ [7]. To resolve this puzzling difference, we measured resistivity anisotropy on single crystals of $\mathrm{Ba}_{0.67}\mathrm{K}_{0.33}\mathrm{Fe}_2\mathrm{As}_2$ and $\mathrm{BaFe}_{1.904}\mathrm{Ni}_{0.096}\mathrm{As}_2$ using the Montgomery method with a mechanical clamp that can vary the applied pressure *in-situ* as described in Ref. [8]. Prefactors due to samples not being perfect squares were corrected for at 200 K [8], where no/negligible resistivity anisotropy is present. Several nominal pressures were applied and the actual zero pressure is determined by performing a linear fit of the change in resistivity anisotropy as a function of nominal pressure. Measured resistivity anisotropy is then scaled to $P = 15$ MPa for both samples to allow for direct comparison under the same applied pressure. Resistivity anisotropy is linearly proportional to the applied pressure in the pressure range we studied, as shown in Supplementary Figure 2 for $\mathrm{Ba}_{0.67}\mathrm{K}_{0.33}\mathrm{Fe}_2\mathrm{As}_2$, similar to $\mathrm{BaFe}_2\mathrm{As}_2$ in which linear response persists up to 90 MPa in the paramagnetic state [8].

Our results are shown in Fig. 1(c), resistivity anisotropy with reversed sign is observed in optimally electron- and hole-doped $\mathrm{BaFe}_2\mathrm{As}_2$ samples. Significant anisotropy is seen in optimal-doped $\mathrm{Ba}_{0.67}\mathrm{K}_{0.33}\mathrm{Fe}_2\mathrm{As}_2$, in agreement with elastoresistance measurements. The much smaller resistivity anisotropy in $\mathrm{Ba}_{1-x}\mathrm{K}_x\mathrm{Fe}_2\mathrm{As}_2$ and small applied pressure are likely causes of why it was not observed in previous work [6].

### Anisotropy of the spin resonance with spin-orbit coupling

The spin excitations in the pnictides become anisotropic due to the spin-orbit coupling (SOC). At low energies, the excitations are peaked at the ordering vectors $\mathbf{Q}_1 = (\pi, 0)$ and $\mathbf{Q}_2 = (0, \pi)$. Therefore, we introduce two vector

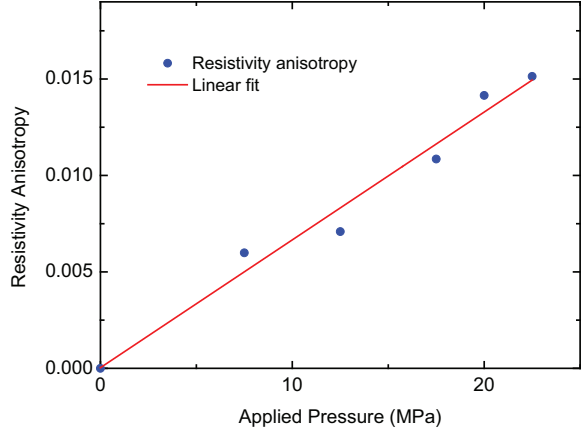

Supplementary Figure 2: Resistivity anisotropy as a function of applied pressure. Data is obtained by combining measured points with $50 \leq T \leq 60$ K, a scaling factor used to normalize resistivity anisotropy to $P = 15$ MPa is obtained by a linear fit as shown.

order parameters, $\mathbf{M}_1$ and $\mathbf{M}_2$. Following the theoretical results of Ref. [9], at sufficient hole-doping concentrations the $c$-axis becomes the easy axis, in agreement with experiments in Na-doped Ba122 [10]. Thus, the quadratic part of the free energy acquires the anisotropic contribution:

$$F(\mathbf{M}_1, \mathbf{M}_2) = \alpha_a \left( M_{1,a}^2 + M_{2,b}^2 \right) + \alpha_b \left( M_{1,b}^2 + M_{2,a}^2 \right) + \alpha_c \left( M_{1,c}^2 + M_{2,c}^2 \right) \tag{S1}$$

To simplify our analysis, hereafter we consider $\alpha_a = \alpha_b \equiv \alpha_{ab}$. According to Ref. [9], $\alpha_c < \alpha_{ab}$. In addition to the reorientation of the moments in the magnetic phase, the anisotropy of the spin excitations also affects the energy of the resonance mode in the $s^{+-}$ superconducting (SC) phase. The spin resonance mode emerges as a spin-1 collective mode protected by the SC gap.

A full calculation of the spin resonance mode in the presence of spin-orbit coupling is beyond the scope of this paper (see Ref. [11]); here, we illustrate the effect of the spin anisotropy on the resonance mode using a semi-phenomenological approach. We consider a simple two band model containing one hole and one electron pocket. For simplicity, we assume perfect nesting – deviations from perfect nesting may shift the wave-vector of the resonance mode, as explained in Ref [12]. At low temperatures and deep inside the SC dome, we can treat the SC gap $\Delta$ as a constant. For a system without spin anisotropy, and within RPA, the resonance energy $\omega$ in the $s^{+-}$ SC phase is given by the condition $\chi^{-1}(\mathbf{Q},\omega) = U/2$, where $U$ is the density-density interaction projected in the spin-spin channel, and $\chi^{-1}(\mathbf{Q},\omega)$ is the isotropic non-interacting susceptibility inside the SC state. Performing the calculation, we find an implicit equation for $\omega$:

$$\frac{1}{2UN_f} = \ln\frac{2\Lambda}{\Delta} + \frac{\omega}{\sqrt{(2\Delta)^2 - \omega^2}} \tan^{-1}\frac{\omega}{\sqrt{(2\Delta)^2 - \omega^2}} \;, \tag{S2}$$

Here, $\Lambda$ is the high energy cutoff and $N_f$ is the density of states at the Fermi level. For small $U$, $\omega \lesssim 2\Delta$. As $U$ increases, a magnetically ordered state appears inside the SC dome at the critical value $U^* = \left[2N_f \ln\left(\frac{2\Lambda}{\Delta}\right)\right]^{-1}$, and the resonance mode vanishes.

Phenomenologically, the main effect of the spin anisotropy, as shown by Eq. (S1), is to shift the different components of the non-interacting spin susceptibility according to

$$\chi^{-1}(\mathbf{Q},\omega) \to \chi_{ii}^{-1}(\mathbf{Q},\omega) = \chi^{-1}(\mathbf{Q},\omega) + \alpha_i \quad i = ab \,, \; c \tag{S3}$$

As a result, the RPA condition $\chi_{ii}^{-1}(\mathbf{Q},\omega) = U/2$ will give different resonance energies $\omega_i$ for different channels. Effectively, the critical value of the interaction $U^*$ depends on the polarization channel. Since $\alpha_c < \alpha_{ab}$, we have $U_c^* < U_{ab}^*$. Physically, this means that for the same value of $U$, $M_c$ is closer to its instability than $M_a$ or $M_b$. Evaluation of $\omega_i$ then gives (assuming $\alpha_i$ small):

$$\frac{\omega_i}{\sqrt{(2\Delta)^2 - \omega_i^2}} \tan^{-1}\frac{\omega_i}{\sqrt{(2\Delta)^2 - \omega_i^2}} = \frac{1}{2N_f} \left( \frac{1}{U} - \frac{1}{U_i^*} \right) \;, \tag{S4}$$

Since $U_c^* < U_{ab}^*$, $\omega_c < \omega_{ab}$ in general, as shown in Fig. 4(c). The difference increases as $U_c^*$ is approached, since at that point $\omega_c$ vanishes but $\omega_{ab}$ remains finite. For that figure, we used the parameters $U_{ab}^* N_f = 4/19$ and $U_c^* N_f = 0.2$.

---



\end{document}